\documentclass[pdflatex,sn-nature]{sn-jnl} % Style for submissions to Nature Portfolio journals
\usepackage{graphicx}%
\usepackage{multirow}%
\usepackage{amsmath,amssymb,amsfonts}%
\usepackage{amsthm}%
\usepackage{mathrsfs}%
\usepackage[title]{appendix}%
\usepackage{xcolor}%
\usepackage{textcomp}%
\usepackage{manyfoot}%
\usepackage{booktabs}%
\usepackage{algorithm}%
\usepackage{algorithmicx}%
\usepackage{algpseudocode}%
\usepackage{listings}%
\usepackage{hyperref}
\usepackage[normalem]{ulem}
\let\orcidlogo\undefined
\usepackage{orcidlink}
\theoremstyle{thmstyleone}%
\theoremstyle{thmstyletwo}%

\theoremstyle{thmstylethree}%

\newcommand{\ve}[1]{\boldsymbol{#1}}
\renewcommand{\vec}[1]{\boldsymbol{#1}}

\newcommand{\ket}[1]{\left| #1 \right>}
\newcommand{\bra}[1]{\left< #1 \right|}
\newcommand{\braket}[2]{\left< #1 | #2 \right>}
\newcommand{\ketbra}[2]{\left| #1 \right> \left< #2 \right|}

\renewcommand{\c}{\hat{c}}
\newcommand{\cd}{\hat{c}^\dagger}

\renewcommand{\a}{\hat{a}}
\newcommand{\ad}{\hat{a}^\dagger}

\newcommand{\n}{\hat{n}}

\newcommand{\D}[1]{\mathcal{D}[#1]}

\begin{document}

\title{Probing metal-insulator criticality with cavity photons}


\author*[1]{\fnm{Jo\~ao} \sur{C. In\'acio} \orcidlink{0009-0006-5457-3711}}\email{joao.carvalho-inacio@uni-wuerzburg.de}
\author[1,3]{\fnm{Natanael} \sur{C. Costa} \orcidlink{0000-0003-4285-4672}}
\author[1,2]{\fnm{Fakher} \sur{F. Assaad} \orcidlink{0000-0002-3302-9243}}

\affil*[1]{\orgdiv{Institut f\"ur Theoretische Physik und Astrophysik}, \orgname{Universit\"at W\"urzburg}, \orgaddress{\city{W\"urzburg}, \postcode{97074}, \country{Germany}}}
\affil[2]{\orgname{W\"urzburg-Dresden Cluster of Excellence ctd.qmat}, \orgaddress{\country{Germany}}}
\affil[3]{\orgdiv{Instituto de F\'isica}, \orgname{Universidade Federal do Rio de Janeiro}, \orgaddress{\street{Cx.P. 68.528}, \city{Rio de Janeiro RJ}, \postcode{21941-972}, \country{Brazil}}}

\abstract{ 
  The Hubbard model on the honeycomb lattice is a pristine realisation of a semimetal-to-insulator Mott transition belonging to the O(3) Gross-Neveu universality class. We couple this system to a single linearly polarised, uniform cavity photon mode. For this interacting light-matter model, we formulate a negative-sign-free fermion quantum Monte Carlo algorithm that allows for unbiased results on finite system sizes. Our numerical results show that the coupling to the cavity is irrelevant at criticality, even at strong electron-photon coupling. On the other hand, we observe, and show analytically, that the photon spectrum acquires finite corrections in the form of the optical conductivity. In the semimetal phase, gapless particle-hole excitations produce a low-energy polaritonic mode, which is suppressed in the Mott insulator phase. Furthermore, we show that the photon spectra can be used to detect universal information about the quantum critical point. Thereby, our results establish that a single, uniform cavity photon mode acts as a contact-free non-invasive probe for Mott criticality. Our results can be generalised to  any cavity  where an \textit{intensive} number of uniform linearly polarised photon modes couple to the electronic system.
}

% \keywords{light-matter interaction, Hubbard model, Dirac fermions, quantum Monte Carlo}

\maketitle
% \tableofcontents

\section{Introduction}
\label{sec:introduction}

The possibility of controlling and probing quantum matter through strong or ultrastrong light-matter coupling in cavities has opened a route in which light becomes an active component of the many-body environment \cite{garcia-vidal_manipulating_2021,schlawin_cavity_2022,lu_cavity_2025,mivehvar_cavity_2021,bretscher_fluctuation_2026}.  In contrast to conventional Floquet protocols, where classical light produces transient nonequilibrium states, cavity quantum materials exploit confined quantised photon modes whose vacuum fluctuations can modify and probe equilibrium properties when the electromagnetic mode volume is sufficiently reduced.
This has already been demonstrated in quantum-gas cavity quantum electrodynamics (QED), where cavity fields can tune long-range order \cite{mivehvar_cavity_2021,landig_quantum_2016,helson_density-wave_2023,zwettler_cavity-mediated_2025}.
In this context, recent experiments reporting cavity-induced modifications of both integer and fractional quantum Hall states provide further motivation to ask whether analogous cavity-mediated mechanisms can be realised in solid-state quantum materials \cite{appugliese_breakdown_2022,enkner_testing_2024,enkner_tunable_2025}.

Two-dimensional compounds constitute a particularly compelling platform for addressing this question, with some experimental progress for graphene-based or van der Waals materials. Experiments in optical microcavities have shown that cavity confinement can control photocurrent generation and transport properties in graphene \cite{engel_lightmatter_2012}, while time-resolved photoemission experiments have reported light-induced anomalous Hall responses and Floquet-dressed Dirac bands \cite{mciver_light-induced_2020,merboldt_observation_2025,choi_observation_2025}. In addition, cavity studies of Bernal bilayer graphene further suggest that confined quantum fields can generate attractive interactions \cite{helmrich_cavity-driven_2026,keren_cavity-altered_2026}. Experiments on the charge-density-wave compound $1T$-$\mathrm{TaS}_{2}$ embedded in tunable THz cavities have further shown a reversible shift of its metal-to-insulator transition, suggesting that cavity QED can modify the thermodynamic environment and macroscopic transport properties of correlated quantum materials \cite{jarc_cavity-mediated_2023}.

Indeed, one of the central theoretical promises of cavity quantum materials is the possibility of engineering cavity-mediated interactions, beyond single-particle band dressing \cite{schlawin_cavity_2022,lu_cavity_2025}.
In lattice models, the Peierls coupling to a quantised vector potential dresses electronic hopping by photon operators and thereby generates photon-assisted processes that are absent in the bare Hamiltonian \cite{dmytruk_gauge_2021,li_effective_2022}.  In correlated systems, such processes can renormalise exchange interactions, induce long-range spin couplings, and generate density-density interactions mediated by virtual or driven cavity photons \cite{fadler_engineering_2024,wang_excitonic_2024}.  
Interestingly, Refs.~\cite{nakamoto_one-dimensional_2025,kass_many-body_2024,grunwald_cavity_2025,sur_amplified_2025,passetti_cavity_2023} found that the photon field may encode the spin and charge correlations, suggesting a new route to examine the correlations in many-body systems \cite{schlawin_cavity_2022,lu_cavity_2025,sur_amplified_2025}.
This perspective is particularly appealing for metal-insulator transitions, where the reconstruction of low-energy charge excitations can leave signatures in the cavity spectrum. A central question is whether these signatures reveal not only the opening of an insulating gap but also universal information about the underlying quantum critical point, while leaving its critical behaviour unchanged.

Ideal Fabry-P\'erot cavities consisting of two parallel mirrors provide a natural setting for addressing these questions. 
The distance between the two mirrors leads to a massive  photon dispersion with quantised transverse momenta. For small distances, only a few transverse momenta can  be taken into account. The wavelength of THz light, $10^{-4}\ \mathrm m$, greatly exceeds the lattice constant of quantum materials set  by  $10^{-10}\ \mathrm m$. It is hence an excellent approximation to set  the longitudinal wave vector of the cavity modes  to zero  \cite{jackson1999}. In other words the, cavity mode does not provide a source of momentum transfer  for scattering processes in the quantum material. 
We focus on the limit in which a fixed number of spatially uniform modes, all  with real polarisation vectors, couple to the matter, with the vacuum-field amplitude of each mode decreasing as the inverse square root of the sample area. 
The number of coupled modes then remains intensive as the number of electronic degrees of freedom grows; a single mode is the simplest representative of this limit. 
Here, we investigate this setting using the half-filled honeycomb Hubbard model coupled to a uniform cavity mode through exact auxiliary-field quantum Monte Carlo simulations. 
Our results are consistent with unchanged Gross-Neveu O(3) critical behaviour, while the photon spectrum retains a finite contribution governed by the electronic optical conductivity. This establishes a route to detecting the semimetal-to-Mott-insulator transition through cavity spectroscopy and predicts access to universal information encoded in the critical conductivity.

The  paper is organised as follows. In Sec.~\ref{sec:results}, we present our main results, including the behaviour of the electronic and photonic sectors. Sec.~\ref{sec:conclusion} provides a discussion of the implications of our findings. Finally, Sec.~\ref{sec:methods} details the methods used in our study.

\begin{figure}[t]
  \centering
  \includegraphics[width=.75\textwidth]{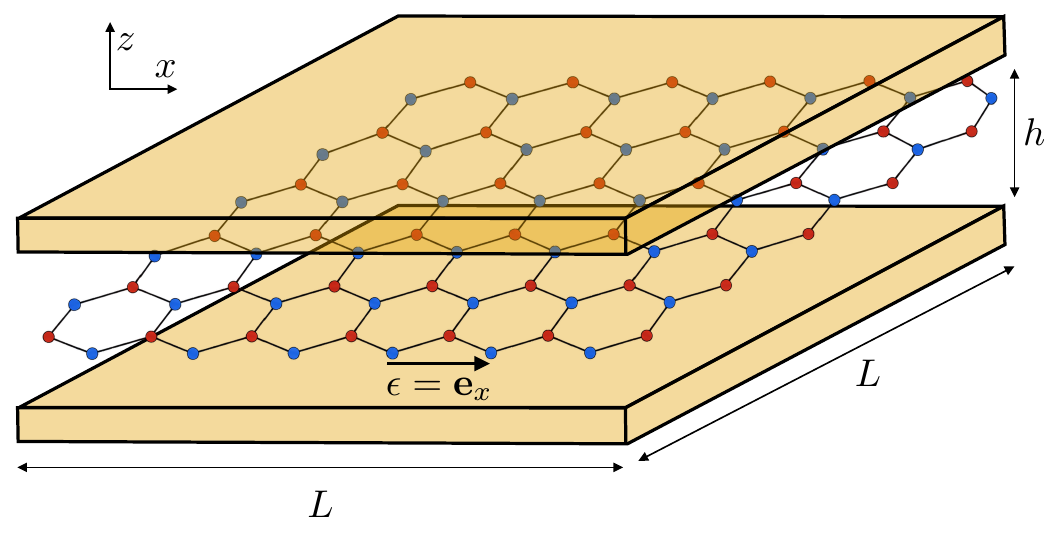}
  \caption{\textbf{Schematic cavity setup.} A half-filled honeycomb Hubbard layer between parallel cavity mirrors couples to a linearly polarised, spatially uniform (\(\vec{q}=0\)) mode with \(\vec{\epsilon}=(1,0)\).}
  \label{fig:schematic_phase_diagram}
\end{figure}

\section{Results}
\label{sec:results}

\subsection{Cavity-coupled honeycomb Hubbard model}

We consider the half-filled Hubbard model on the honeycomb lattice as a minimal setting for studying the influence of photon vacuum fluctuations on a quantum critical point \cite{herbut_interactions_2006,assaad_pinning_2013,wang_resolving_2026}. The electronic system is placed inside an optical cavity formed by two parallel plates of area \(A\) separated by a distance \(h\), as depicted in Fig.~\ref{fig:schematic_phase_diagram}. 
In an ideal Fabry-P\'erot cavity, the confined classical electric field is given by 
\begin{equation}
  \ve{E}(\ve{x},t)=    \text{Re} \sum_{n}  \int d^2 \ve{q}_{\parallel}   \ve{E}_{n}(\ve{q}_{\parallel}) e^{i (\ve{q}_{\parallel} \cdot \ve{x}_{\parallel} - \Omega_n(\ve{q}_{\parallel}) t)}   \sin\left( \frac{n \pi z}{h} \right),
\end{equation}
where the  dispersion relation is  given by \(\Omega_n(\ve{q}_{\parallel}) = \sqrt{\ve{q}_{\parallel}^2 + (n \pi / h)^2}\). We assume linear polarisation for each mode, \(\ve{E}_n(\ve{q}_{\parallel})\in\mathbb{R}^3\), take \(\ve{q}_{\parallel}\to0\) as argued above, retain only a finite number of modes, and set \(z=h/2\) without loss of generality. Setting \(\hbar = c = e = 1\), the  corresponding quantised vector potential is  \cite{loudon_quantum_2010}
\begin{equation}
  \vec{\hat{A}} = \sum_{n=1}^{N_{\mathrm{modes}}} \sqrt{\frac{1}{2 \epsilon_0 V \Omega_n}} \vec{\epsilon}_n (\a_n + \ad_n),
\end{equation}
where \(V = Ah\) is the cavity volume. Coupling to the electronic degrees of freedom is achieved via the Peierls substitution \cite{dmytruk_gauge_2021}. Since our vector potential has no spatial dependence \(\nabla \times \vec{\hat{A}} = 0\), the induced magnetic field is zero; thus, no Zeeman-like coupling is generated. The resulting light-matter Hamiltonian is
\begin{multline} \label{eq:hamiltonian}
  \hat{H} = - t \sum_{i \in A, \delta, \sigma} \left(\cd_{i, \sigma} \prod_{n=1}^{N_{\mathrm{modes}}} e^{i \frac{g_n}{\sqrt{N}} \vec{\epsilon}_n \cdot \vec{n}_\delta \hat{X}_n} \hat{c}_{i+\delta, \sigma} + \text{h.c.}\right) + \\
  \frac{U}{2} \sum_i(\n_i-1)^2 + \sum_{n=1}^{N_{\mathrm{modes}}} \frac{\Omega_n}{2} (\hat{P}_n^2 + \hat{X}_n^2),
\end{multline}
where \(i\) labels a site on sublattice \(A\) in unit cell \(\vec{r}_i\) on the honeycomb lattice with \(N = L^2\) unit cells. The nearest-neighbour vectors \(\vec{n}_\delta\) are index by \(\delta=1,2,3\), and \(\sigma =\ \uparrow,\downarrow\) is the spin index. Furthermore, \(\{\hat{c}_{i,\sigma}, \cd_{j,\sigma^\prime}\} = \delta_{i,j}\delta_{\sigma,\sigma^\prime}\) and \(\hat{n}_i = \sum_\sigma \cd_{i, \sigma} \hat{c}_{i, \sigma}\). The dimensionless photon operators \(\hat{X}_n\) and \(\hat{P}_n\) specify the position and momentum of the \(n\)-th photon mode, respectively, with \([\hat{X}_n, \hat{P}_m] = i\delta_{n,m}\). The relation to the creation and annihilation operators is given by \(\hat{X}_n = (\ad_n + \a_n) / \sqrt{2}\), \(\hat{P}_n = i (\ad_n - \a_n) / \sqrt{2}\), where \([\a_n, \ad_m] = \delta_{n,m}\) and, after dropping the constant zero-point energy, the bare photon Hamiltonian is given by \(\hat{H}_{\text{ph}} = \sum_{n=1}^{N_{\mathrm{modes}}} \Omega_n \ad_n\a_n\).

To obtain a well-defined thermodynamic limit, we scale the cavity area with the number of unit cells, \(A \sim N\), known as the intensive light-matter coupling. The effective light-matter coupling is given as \cite{ashida_nonperturbative_2022-1}:
\begin{equation}
  g_{\mathrm{eff}}^2 = \sum_{n=1}^{N_{\mathrm{modes}}} \frac{g_n^2}{N} \sim \frac{N_{\mathrm{modes}}}{N}.
\end{equation}
The \(1/N\) factor ensures that the photon occupation remains intensive as the system size increases. Moreover, we consider the physical limit of few modes, i.e. \(g_{\mathrm{eff}} \to 0\) in the thermodynamic limit. In our quantum Monte Carlo calculations we consider the simplest case of this limit: the single-mode limit, i.e. \(N_{\mathrm{modes}} = 1\), while retaining our analytical predictions for a non-extensive number of modes. The single mode has a light-matter coupling strength \(g\) and frequency \(\Omega\). With this notation, the dimensionless light-matter coupling is given by \(g = \sqrt{\frac{\pi\alpha}{h \Omega}}\), where \(\alpha\) is the fine-structure constant. In an empty cavity with vanishing in-plane momentum, the vertical boundary conditions give \(\Omega_n = n \pi / h\). Combining this with the expression above gives \(g_n=\sqrt{\alpha/n}\). Consequently, changing \(h\) at fixed mode index \(n\) varies the cavity frequency whilst leaving \(g\) unchanged. Conversely, at fixed \(\Omega\), different modes can be selected by choosing \(h_n = n\pi/\Omega\), yielding again \(g_n = \sqrt{\alpha/n}\). Then, \(g\) and \(\Omega\) are geometrically linked. In our numerical analysis, we consider \(g\) and \(\Omega\) as independent parameters, allowing us to isolate effects arising from the cavity frequency from the coupling strength \cite{eckhardt_quantum_2022,passetti_cavity_2023,nakamoto_one-dimensional_2025}. A microscopic realisation of different \((g, \Omega)\) pairs should therefore be interpreted as representing different engineered cavity environments rather than independent tuning of one fixed ideal cavity. 

The Hamiltonian in Eq.~\eqref{eq:hamiltonian} preserves the global symmetries of the Hubbard model, such as particle-hole (PH), SU(2) spin and U(1) charge symmetries. In addition, the system remains invariant under time-reversal symmetry, since the vector potential is odd under this transformation. Moreover, the model preserves lattice translation symmetry, but the selected polarisation breaks the \(C_{3v}\) point group symmetries of the honeycomb lattice, as detailed in the Supplementary Information. 

To study the light-matter Hamiltonian Eq.~\eqref{eq:hamiltonian}, we use the Algorithms for Lattice Fermions \cite{assaad_alf_2025} implementation of the finite-temperature auxiliary-field quantum Monte Carlo (AF-QMC) method \cite{blankenbecler_monte_1981,white_numerical_1989,assaad_world-line_2008}. Details can be found in the "Methods" section. We use \(\Delta\tau = 0.1\). For all of our results we set \(\Omega = t = 1\). All of the spectral data shown here were obtained using the stochastic analytic continuation method (SAC) \cite{sandvik_stochastic_1998,beach_identifying_2004,shao_progress_2023} implemented in the ALF package on the imaginary-time correlation functions measured in the AF-QMC simulations.

\subsection{Criticality and electronic spectral function}

Without coupling to photons, the Hubbard model on the honeycomb lattice has a quantum critical point (QCP) between a Dirac semimetal (DSM) phase and a Mott-insulator phase with antiferromagnetic (AFM) long-range order. The critical point belongs to the O(3) Gross-Neveu (GN) universality class \cite{herbut_interactions_2006,assaad_pinning_2013,wang_resolving_2026}. Our primary objective is to determine whether the GN critical point is stable upon coupling to photons. To characterise the Mott transition, we measure the AFM correlation ratio 
\begin{equation} \label{eq:correlation_ratio}
  r_{\text{AFM}} = 1 - \frac{S^{\text{S}}(\vec{\Gamma} + \delta\vec{q})}{S^{\text{S}}(\vec{\Gamma})}
\end{equation}
where \(S^{\text{S}}(\vec{q}) = \frac{1}{N} \sum_{i,j} e^{i\vec{q}\cdot(\vec{r}_i - \vec{r}_j)} (\langle \vec{\hat{S}}_{i} \cdot \vec{\hat{S}}_{j} \rangle - \langle \vec{\hat{S}}_{i} \rangle \cdot \langle \vec{\hat{S}}_{j} \rangle)\) is the spin structure factor, with \(\vec{\hat{S}}_{i} = \vec{\hat{S}}_{i, A} - \vec{\hat{S}}_{i, B}\). The AFM ordering wave vector is \(\vec{\Gamma}\) (\(\vec{\Gamma}'\) in the unfolded Brillouin zone) and \(|\delta \vec{q}| \sim 1 / L\) is the smallest wave vector away from \(\vec{\Gamma}\). In the thermodynamic limit, \(r_{\text{AFM}}\) converges to zero (unity) in a disordered (ordered) phase. Furthermore, it is an RG-invariant quantity, such that at a critical point we expect \(r_{\text{AFM}}(L, U) = f(L^z/\beta, (U - U_c)L^{1/\nu}, L^{-\omega})\), where \(z\) is the dynamical exponent, \(U_c\) is the value of the critical coupling, \(\nu\) is the correlation-length exponent and \(\omega\) is the leading correction to the scaling exponent. Due to our small sample of system sizes, we ignore the correction term. At a QCP, Lorentz symmetry is emergent, which implies \(z=1\). Thus, when we set \(\beta=L\), the \(r_{\text{AFM}}\) curves of different system sizes cross at the critical point.

\begin{figure}[t]
  % g = 0.5
  % 0.04383369582544409 0.0007470567747856194
  % 0.41736264808834794 0.0004426436383568674
  % g = 1.0
  % 0.04481527981685158 0.0006430829561138807
  % 0.4199383242407159 0.0008891492273140418
  % g = 2.0
  % 0.04393887687666085 0.0007831217572136652
  % 0.4253066231713582 0.0009186677551037866
  % g = 4.0
  % 0.045409463347727566 0.0004816285090819401
  % 0.4407321782771083 0.0004022214054492059
  \centering
  \begin{tabular}{cc}
    (a) & (b) \\
    \includegraphics[width=.45\textwidth]{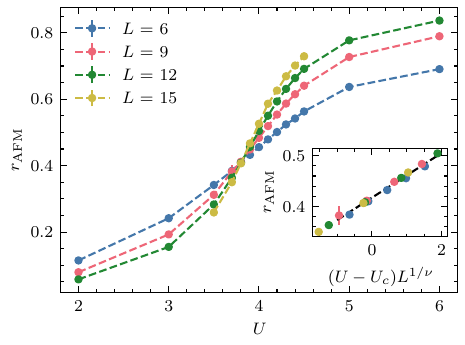} &
    \includegraphics[width=.45\textwidth]{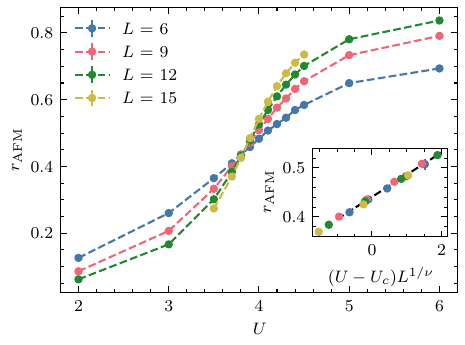}
  \end{tabular}
  \caption{
    \textbf{Finite-size scaling of the antiferromagnetic correlation ratio.}
    The antiferromagnetic correlation ratio \(r_{\mathrm{AFM}}\) as a function of \(U\) at \textbf{a}, \(g = 0.5\) and \textbf{b}, \(g = 4\). The insets show \(r_{\mathrm{AFM}}\) versus \((U-U_c)L^{1/\nu}\), using \(U_c = 3.819(3)\) and \(\nu = 1.06(2)\) from Ref.~\cite{wang_resolving_2026}. Black dashed lines are linear fits, yielding \(y=0.0438(7)x+0.4174(4)\) in \textbf{a} and \(y=0.0454(5)x+0.4407(4)\) in \textbf{b}.
  }
  \label{fig:rAFM_collapse}
\end{figure}

Figure \ref{fig:rAFM_collapse} shows the AFM correlation ratio as a function of \(U\) and \(L\) for \(g = 0.5, 4\), respectively. In both cases, the onset of AFM order occurs at approximately the same critical interaction strength \(U_c\). To investigate whether the transition remains in the GN universality class, we perform a data collapse (see the insets of Fig.~\ref{fig:rAFM_collapse}). For the scaling analysis, we use the critical parameters for the GN transition taken from Ref.~\cite{wang_resolving_2026}, \(U_c = 3.819(3)\) and \(\nu = 1.06(2)\). For the accessible system sizes, the crossings remain close to the uncoupled critical interaction. Rescaling the data using established GN critical parameters produces a comparable collapse for both cavity couplings. The results are therefore consistent with unchanged GN critical scaling within our numerical resolution.

To understand why the electron-photon coupling is irrelevant at criticality, we analyse the fermionic channels to which the photon couples. To this end, we write the Hamiltonian \eqref{eq:hamiltonian} in the coherent-state path integral formalism, see "Methods". After a gauge transformation, we can integrate out the photons to obtain an action $S = S_{\text{tU}} + S_{\text{e-ph}}$, where $S_{\text{tU}}$ is the action of the Hubbard model and
\begin{equation} \label{eq:effective_action}
  S_{\text{e-ph}} = - \sum_{n=1}^{N_{\mathrm{modes}}} g_n^2 \int_0^\beta d\tau d\tau^\prime\, J_{\epsilon_n}(\vec{q} = 0, \tau) D_0(n, \tau - \tau^\prime) J_{\epsilon_n}(\vec{q} = 0, \tau^\prime),
\end{equation}
is the effective electron-photon interaction. Here $D_0(n, \tau) = \cosh(\Omega_n(\beta/2 - |\tau|))/\sinh(\beta\Omega_n/2)$ is the bare photon propagator of the \(n\)-th mode, and $J_{\epsilon_n}(\vec{q} = 0, \tau)$ the total current density along the $\epsilon_n$ direction.
Equation \eqref{eq:effective_action} shows that the photons induce a retarded, all-to-all current-current interaction and thus couple exclusively to PH excitations with total momentum $\vec{q} = 0$ and total spin $S = 0$. In particular, the photons probe only uniform charge current fluctuations and do not couple to finite-momentum spin fluctuations. This is fundamentally different from the critical fluctuations at the GN critical point, which are fluctuations of the AFM order parameter that carry spin and generally have non-zero momentum.
Consequently, the photon-mediated interaction does not couple directly to the soft critical modes that control the GN transition.

The mismatch in both momentum and spin quantum numbers strongly suppresses the effect of the photons on the critical physics. However, this does not preclude the generation of new phases or effects at large \(g\). While the action of the Hubbard model, $S_{\text{tU}}$, is \textit{extensive}, \(S_{\text{e-ph}}\) is \textit{intensive}. Therefore, the photonic contribution to the free energy density vanishes in the thermodynamic limit. Thus, even at large \(g_n\), the photons generate at most subleading corrections to electronic properties but will not alter the critical behaviour or induce new electronic phases.

\begin{figure}[t]
  \centering
  \begin{tabular}{cc}
    (a) & (b) \\
    \includegraphics[width=.45\textwidth]{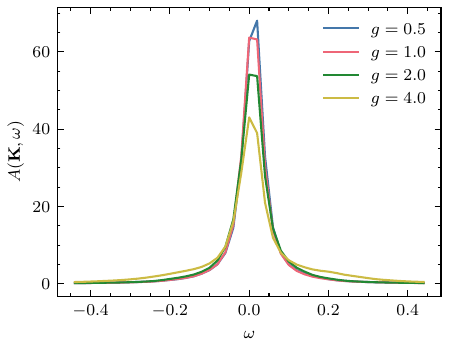} &
    \includegraphics[width=.45\textwidth]{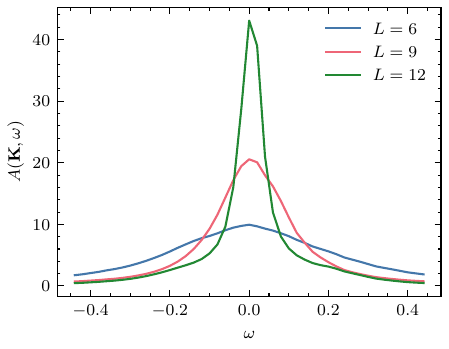}
  \end{tabular}
  \caption{
    \textbf{Electronic spectral function at the Dirac point.}
    \textbf{a}, \(A(\vec{K},\omega)\) for \(g = 0.5, 1, 2\) and \(4\) at \(L = 12\).
    \textbf{b}, \(A(\vec{K},\omega)\) for \(L = 6, 9\) and \(12\) at \(g = 4\). 
    All results are for \(U=2\) in the Dirac-semimetal phase. The spectra are obtained by SAC of \(G(\vec{K}, \tau) = \langle \hat{c}^\dagger_{\vec{K}}(\tau) \hat{c}_{\vec{K}}(0) \rangle\).
  }
  \label{fig:green_fermion_gs}
\end{figure}

The intensive nature of the electron-photon interaction is reflected in the single-particle Green's function \(G(\vec{k}, \omega) = (\omega + i\eta - \epsilon_{\vec{k}} - \Sigma_U(\vec{k}, \omega) - \Sigma_{\text{e-ph}}(\vec{k}, \omega))^{-1}\), where \(\epsilon_{\vec{k}}\) is the non-interacting energy, \(\eta \to 0^+\), and \(\Sigma_U\) and \(\Sigma_{\text{e-ph}}\) are the self-energies associated with the Hubbard and electron-photon interactions, respectively. At fixed \(U\), we can isolate the effects of the electron-photon self-energy on the single-particle Green's function. Within our path-integral formulation, see "Methods", the electron-photon self-energy to lowest order in \(g\) is given by 
\begin{equation}
  \Sigma_{\text{e-ph}}(\vec{k}, i\omega_n) = \sum_{n'} \frac{g_{n'}^2}{N} j_{\vec{k}} \left(\frac{1}{\beta} \sum_m D_0(n', i\Omega_m) G_0(\vec{k}, i\Omega_m + i\omega_n) \right) j_{\vec{k}},
\end{equation}
where \(D_0\) is the photon Green's function, \(G_0\) is the bare single-particle Green's function and \(j_{\vec{k}}\) is the current matrix element. \(\Sigma_{\text{e-ph}}\) explicitly shows a \(1/N\) scaling and therefore the photon-induced correction vanishes in the thermodynamic limit. To verify this behaviour, we examine the spectral function \(A(\vec{K}, \omega) = - 2 \text{Im} G(\vec{K}, \omega)\), at fixed \(U = 2\), as a function of system size and light-matter coupling, Fig.~\ref{fig:green_fermion_gs}. The full momentum-resolved spectra are presented in the Supplementary Information. As \(g\) increases, the quasiparticle peak is progressively suppressed and broadened. Owing to the spectral sum rule, this reduction in peak height is accompanied by a redistribution of spectral weight towards finite frequencies, indicating a reduced quasiparticle lifetime. Moreover, this broadening substantially decreases with increasing system size. This implies that the real part of the self-energy is negligible, while the imaginary part scales as \(\Sigma^{\prime\prime}_{\text{e-ph}} \sim 1/N\). Thus, the electronic spectral function provides clear evidence that the effective electron-photon self-energy corrections are purely finite-size effects and do not modify the electronic structure in the thermodynamic limit.

\subsection{Photonic spectral functions}

Having established that the photons do not alter the physics of the Hubbard model, a natural question is whether the Mott transition can nevertheless be detected through photonic observables. In the DSM phase the fermions possess gapless PH excitations, while in the Mott phase only magnons remain at low energies. Since the cavity photons couple directly to the electronic degrees of freedom, this qualitative change in the low-energy excitation spectrum should be reflected in the photon spectral function \(B(\omega)\). It characterises the frequency-resolved response of the cavity field and determines the spectra measured in transmission, reflection, and emission experiments \cite{zoubi_excitons_2007,mekhov_probing_2007,dendzik_observation_2020}. 

\begin{figure}[t]
  \centering
  \begin{tabular}{cc}
    (a) & (b) \\ 
    \includegraphics[width=.45\textwidth]{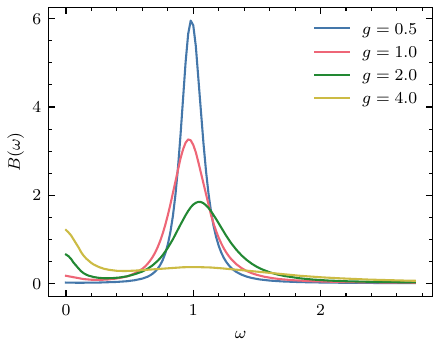} &
    \includegraphics[width=.45\textwidth]{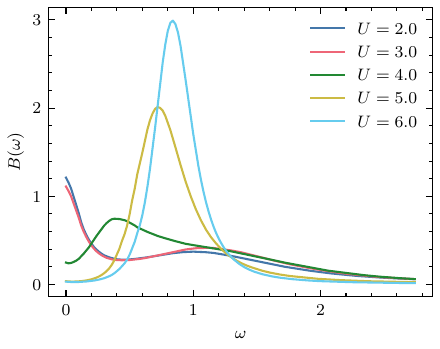}
  \end{tabular}
  \caption{
    \textbf{Photon spectral function across the interaction-driven transition.}
    \textbf{a}, \(B(\omega)\) for \(g=0.5, 1, 2\) and \(4\) at \(U = 2\).
    \textbf{b}, \(B(\omega)\) for \(U=2, 3, 4, 5\) and \(6\) at \(g = 4\); these values span the O(3) GN transition near \(U_c \simeq 3.82\). 
    All results are for \(L = 12\). The spectra are obtained by SAC of \(D(\tau) = \langle \hat{X}(\tau) \hat{X}(0) \rangle\).
  }
  \label{fig:spectral_gs}
\end{figure}

Figure \ref{fig:spectral_gs} shows the photon spectral function in the ground state as a function of \(g\) for \(U = 2\), panel (a), and as a function of \(U\) for \(g = 4\), panel (b). In the DSM phase, panel (a), at small \(g\) the photon mode undergoes a small renormalisation of its frequency and lifetime. As the electron-photon coupling is increased, a new low-energy mode gradually emerges near \(\omega = 0\). We associate this feature with the emergence of a polariton mode, originating from the hybridisation between photons and the low-energy electronic PH excitations. Importantly, the emergence of this low-energy mode does not signal a superradiant transition as the photons and electrons completely decouple at \(\omega = 0\) \cite{dmytruk_gauge_2021,andolina_cavity_2019,andolina_no-go_2021}. Instead, the polariton mode reflects the strong coupling between photons and gapless PH excitations. Upon increasing \(U\), panel (b), the polariton mode is progressively suppressed and eventually disappears as the system enters the Mott-insulator phase. In this phase, charge excitations are gapped and the only remaining low-energy gapless excitations are the spin waves of the AFM state. As these collective modes carry total spin \(S = 1\), they do not couple to the photons according to the selection rules discussed above. Thus, the photon spectrum loses its low-energy polaritonic features in the Mott phase. Therefore, the photon spectral function directly inherits the signatures of the electronic Mott transition, allowing the latter to be detected through purely photonic observables.

Within our path-integral formulation we can \textit{exactly} calculate the electronic contributions to the photon spectral function, see "Methods". One obtains:
\begin{equation} \label{eq:photon_spectral_function}
  D(\omega) = \sum_n D_0(n, \omega)  + i \omega  \sum_n \left[ g_n^2 D_0^2(n, \omega) \right] \sigma^{\text{reg}}_{\epsilon_n,\epsilon_n} (\ve{q}=0,\omega)
\end{equation}
where \(\sigma^{\text{reg}}_{\epsilon,\epsilon}\) is the conductivity of the model in Eq.~\eqref{eq:hamiltonian} along the \(\epsilon\) direction. For $\omega \neq \Omega_n$ the above reduces to
\begin{equation} \label{eq:photon_spectral_function_low_omega}
  B(\omega) = 2 \sum_n \left[ g_n^2 \frac{\Omega_n^2}{(\omega^2 - \Omega_n^2)^2} \right] \omega \text{Re} \sigma^{\text{reg}}_{\epsilon_n,\epsilon_n} (\ve{q}=0,\omega),
\end{equation}
where \(B(\omega) = -2 \text{Im} D(\omega)\) denotes the  photon spectral function. This expression directly shows that the photons acquire a finite self-energy correction, as they pick up the optical conductivity. Since the spectral weight of the polariton mode scales as \(g^2\), it is strongly suppressed at weak coupling \(g < 1\), making it difficult to resolve numerically. In contrast to the electronic spectral function, where photon corrections vanish in the thermodynamic limit, the photon spectral function acquires well-defined corrections in the thermodynamic limit in the form of the conductivity, as the photon field itself is intensive.

The emergence of the polariton mode can then be understood from the low-frequency behaviour of the conductivity, namely the dc limit \(\sigma^{\text{dc}} = \text{Re} \sigma(\omega \to 0)\). In the DSM phase, \(\sigma^{\text{dc}}\) is finite due to the presence of gapless Dirac quasiparticles. Consequently, the photons hybridise with the low-energy PH excitations leading to the emergence of the polariton. Conversely, in the Mott-insulator phase, the opening of a charge gap suppresses the low-frequency conductivity, leading to a reduction of the polaritonic spectral weight. As such, the presence or absence of the polariton mode directly reflects the low-energy charge transport properties of the underlying electronic state. Additionally, one can also detect the upper Hubbard band in the photon spectrum \cite{kiffner_mott_2019} however, since \(\Omega \ll U\), its contribution is severely suppressed, see Eq.~\eqref{eq:photon_spectral_function}. 

We have shown that the cavity can detect the opening of a charge gap. However, this does not yet establish that the photon spectrum can capture universal critical behaviour of the electrons. Near the critical point on the DSM side, the zero-temperature conductivity follows the universal relation: \(\text{Re} \sigma^{\text{reg}}(\omega, U) = \frac{1}{2\pi} H\left(\frac{\omega}{\delta}\right)\) \cite{herbut_theory_2009}, where \(\delta \sim |U - U_c|^{z\nu}\) and \(H\) is a universal crossover function. At criticality, \(\delta\) vanishes and \(H(\omega/\delta) \to H(\infty)\) at any finite frequency. Thus, the critical conductivity becomes frequency independent and approaches the universal value of \(\text{Re} \sigma^{\text{reg}}(\omega, U_c) = H(\infty)/(2\pi)\). Using Eq.~\eqref{eq:photon_spectral_function}, the electronic contribution to the photonic spectral function, for \(\omega\ll\Omega_n\), satisfies 
\begin{equation} \label{eq:low-freq_cavity_spectra}
  \frac{B(\omega, U)}{\omega} = \sum_n \frac{g_n^2}{\pi \Omega_n^2} H\left(\frac{\omega}{\delta}\right).
\end{equation}
Here, we have neglected the $\vec{\epsilon}_n$  dependence of the optical conductivity, since at criticality emergent Lorentz symmetry guarantees an isotropic conductivity. At criticality, the cavity spectrum becomes linear in frequency with the slope given by the universal conductivity: \(\sum_n g_n^2H(\infty) / (\pi \Omega_n^2)\). Therefore, the GN criticality can be detected through the data collapse of the cavity spectral function: \(B(\omega, U)/\omega = f(\omega |U - U_c|^{-z\nu})\). Our SAC data does not have sufficient low-frequency resolution to qualitatively capture this behaviour, making this a prediction for frequency-resolved cavity spectroscopy measurements.

\section{Discussion}
\label{sec:conclusion}

% Summary of work 
We studied the interplay between strong electron interactions and cavity QED in the honeycomb Hubbard model coupled to a non-extensive number of linearly polarised photon modes. By this we mean that \(N_{\mathrm{modes}}/N\) is a small parameter in our study. Our definition of the model, and especially the scaling of the light-matter coupling, ensures that the photon number remains finite, as for the empty cavity, even when an insulating material is placed inside the cavity. Through unbiased AF-QMC simulations in the representative single-mode limit, we have investigated the effect of the light-matter coupling on the O(3) GN critical point and shown that the photons are an irrelevant perturbation, even at strong coupling. This robustness stems from the fact that the photons couple exclusively to uniform, spin-singlet current fluctuations and do not couple directly to the critical magnonic modes. More importantly, upon integrating out the photons, the resulting effective interaction is intensive such that its self-energy contribution vanishes in the thermodynamic limit. Therefore, in the limit of a non-extensive number of modes with linear polarisation, the cavity environment cannot change the properties of the underlying electronic system.

By contrast, the photon spectrum acquires finite corrections, even in the thermodynamic limit, through its coupling to the optical conductivity. In the DSM phase, the low-frequency conductivity is finite such that it generates a low-energy mode, which we identify as a polariton. Conversely, in the Mott phase, the opening of an AFM gap suppresses the conductivity and no polaritonic mode is observed. We interpret this behaviour as a "Mott transition of photons". This terminology does not imply an independent bosonic Mott transition or superradiant instability \cite{dmytruk_gauge_2021,andolina_cavity_2019,andolina_no-go_2021}; it describes the disappearance of a low-energy mode in the photon spectra supported by gapless PH electronic excitations. Moreover, at the GN critical point, the conductivity follows a universal relation \cite{herbut_theory_2009} for which the photon spectra can pick up universal information. As the polaritonic spectral weight scales with the number of modes, the inclusion of more modes will enhance this effect, see Eq.~\eqref{eq:photon_spectral_function_low_omega}. Thus, our results establish that a few-mode cavity with vanishing transverse momenta acts as a non-invasive probe of metal-to-insulator transitions and GN criticality.

% Analogy to defects
The notion that a cavity,  with a non-extensive number of photon modes, cannot  change  the   quantum state but  can characterise it is very similar to impurity physics.  Take, for example, a single magnetic impurity embedded in  a metallic host. The single impurity will not alter the thermodynamic properties of the metal. However, below the Kondo temperature, the  impurity spin will be entangled with the metal in the form of a Kondo resonance with very well understood physics in the simple case of a Fermi liquid. Hence the impurity allows us to uncover aspects of the metallic host  without modifying its thermodynamic properties. In particular, Ref.~\cite{Qin26} shows that the interference between two impurities can reveal altermagnetic properties of the metallic  state.

% Experimental parameters 
Relating the numerical parameters to experiments requires some care. The choice of \(\Omega = t\) is a numerical convenience, see Fig.~\ref{fig:autocorrelation}, rather than an experimental requirement. In fact, for a THz cavity, \(\Omega\) is several orders of magnitude lower than \(t\). For any finite \(\Omega\), the self-energy generated by the photon is suppressed as \(1/N\), such that the value of \(\Omega\) does not affect the physics in the thermodynamic limit. For spectroscopy, \(\Omega\) needs to be sufficiently larger than the emergent polariton frequency to avoid overlap with the bare cavity mode. Choosing it too large would also decrease the polariton signal, see Eq.~\eqref{eq:low-freq_cavity_spectra}. Likewise, the simulated light-matter coupling range \(0.5 \leq g \leq 4\) exceeds realistic couplings \cite{lu_cavity_2025,mueller_ultrastrong_2025,frisk_kockum_ultrastrong_2019}. Nevertheless, the polaritonic contribution persists at weaker coupling, but as its spectral weight scales as \(g^2\) its detection becomes harder. In practice, Fabry-P\'erot cavities have more than one mode such that the polariton weight will be enhanced, even at small \(g\), see Eq.~\eqref{eq:photon_spectral_function_low_omega}. Here we have again assumed that an intensive number of modes couple to the quantum  system. Nevertheless, practically one needs sufficient spectral resolution above the empty-cavity background and extrinsic losses. Thus, our predictions hold in experimental ranges of light-matter coupling and photon frequency.

% Moire materials and measurement of universal conductivity 
Moir\'e Dirac materials provide a natural route to GN criticality: near suitable twist angles, the reduced Dirac bandwidth enhances interactions and generates a mass gap, while increasing the bandwidth by varying twist angle or pressure can continuously close the gap and restore a DSM.
Twist-angle tuning is predicted to realise O(2) and O(3) GN criticality between a DSM and a Kramers intervalley-coherent insulator in twisted bilayer graphene \cite{biedermann_twist-tuned_2025,huang_angle-tuned_2025,hawashin_gross-neveu-xy_2025}, and an AFM insulator in twisted \(\mathrm{WSe_2}\) tetralayers \cite{biedermann_dirac_2026,ma_relativistic_2025,li_continuous_2021,xia_bandwidth-tuned_2026}, respectively. Since the transition is inferred by comparing non-universal quantities of samples with different twist angles, its precise location and universality remain difficult to establish. 
A Fabry-P\'erot cavity could provide a complementary contact-free probe of GN criticality in moir\'e materials. The critical point can be located from the emergence of the polariton mode and a scaling collapse of the low-frequency cavity spectra against \(\omega / |\lambda - \lambda_c|^{z\nu}\), where \(\lambda\) denotes the twist angle or displacement field. This also offers a possibility of measuring the universal conductivity \cite{herbut_theory_2009,roy_collisionless_2018}. At criticality and for \(\omega \ll \Omega\), the universal conductivity may be extracted from the slope of the cavity spectra, see Eq.~\eqref{eq:low-freq_cavity_spectra}.

% Kondo breakdown QCP
Moreover, it would be of particular interest to use cavity photons to explore orbital-selective Mott transitions \cite{Vojta10} as realised in Kondo destruction transitions \cite{danu_kondo_2020,raczkowski_breakdown_2022,pan_quantum_2025,Si01,Senthil03,Paschen21,Sachdev20}. At this transition, an extensive subset of charge carriers becomes localised, leading to a strong change in the optical conductivity and hence in the cavity photon spectral function.

% Multimode
An important direction for future work is the extension to spatially resolved electromagnetic fields, where the photon sector scales extensively and can transfer finite momentum to the electronic system \cite{wang_superradiant_2026,bacciconi_first-order_2023,bretscher_fluctuation_2026,keren_cavity-altered_2026,eckhardt_surface-mediated_2025}. In this setup, the photon contribution to the self-energy will stay finite in the thermodynamic limit, possibly inducing new quantum critical behaviour, ordered phases, or light-induced collective phenomena. A particularly interesting setup uses hyperbolic van der Waals resonators as their strongly confined modes span a broad range of momenta \cite{ashida_cavity_2023,eckhardt_surface-mediated_2025,andolina_theory_2020}. In this setting, one can spatially modulate non-uniform charge fluctuations on lattice length scales.

\section{Methods} 
\label{sec:methods}

\subsection{Path-integral formulation of light-matter coupled Hamiltonian} 
\label{subsec:path_integral}

Using a fermion coherent-state path-integral representation, we write the action of our model system, Eq.~\eqref{eq:hamiltonian}, as reads: 
\begin{equation}
  S = S_U +  S_{\text{ph}} + \int_0^{\beta} d\tau  \sum_{\ve{i},\ve{j}} c^{\dagger}_{\ve{i}}(\tau) \left( \partial_{\tau} \delta_{\ve{i},\ve{j}} -t_{\ve{i},\ve{j}} e^{i \sum_n \frac{g_n}{\sqrt{N}} X_n(\tau) \left( \ve{j} - \ve{i} \right) \cdot \ve{\epsilon_n}} \right) c_{\ve{j}}(\tau),
\end{equation}
where $S_U = U \int_{0}^{\beta}  d \tau  \sum_{\ve{i}} ( c^{\dagger}_{\ve{i}}(\tau) c_{\ve{i}}(\tau) - 1 )^2$ and 
$S_{\text{ph}} =  \int_{0}^{\beta} d\tau \frac{1}{2} \sum_n \left( \frac{\dot{X}_n^2(\tau)}{\Omega_n} + \Omega_n X_n^2(\tau) \right)$.  In the above, the 
sum over the spin index is implicit.
We adopt periodic boundary conditions for the fermionic fields 
\begin{equation}
  c_{\ve{i} + \ve{L}}(\tau) = c_{\ve{i}}(\tau),
\end{equation} 
where $\ve{L}$ denotes the linear size of the system.  To proceed, let us carry out the canonical transformation: 
\begin{equation}
  \eta_{\ve{j}}(\tau) = e^{i \sum_n \frac{g_n}{\sqrt{N}} X_n(\tau) \ve{j} \cdot \ve{\epsilon}_n} c_{\ve{j}}(\tau). 
\end{equation}
This transformation leaves the measure of the path integral invariant and transforms the action into:
\begin{equation}
  S = S_t + S_U + S_{\text{ph}} + \int_0^{\beta} d\tau  \sum_{\ve{i}} \eta^{\dagger}_{\ve{i}}(\tau) \left(  - i \sum_n \frac{g_n}{\sqrt{N}} \dot{X}_n(\tau) \ve{i} \cdot \ve{\epsilon}_n  \right) \eta_{\ve{i}}(\tau),
\end{equation} 
where $S_t= \int_0^{\beta} d\tau  \sum_{\ve{i},\ve{j}} \eta^{\dagger}_{\ve{i}}(\tau) \left[ \partial_{\tau} \delta_{\ve{i},\ve{j}} - t_{\ve{i},\ve{j}} \right] \eta_{\ve{j}}(\tau) $.
The Hubbard term remains invariant under the transformation, and the boundary conditions of the fermion fields acquire a time-dependent twist:
\begin{equation}
  \eta_{\ve{i} + \ve{L},\sigma}(\tau) = e^{i \sum_n \frac{g_n}{\sqrt{N}} X_n(\tau) \ve{\epsilon}_n \cdot \ve{L}} \eta_{\ve{i}}(\tau). 
\end{equation}
We can now carry out a partial integration in the time variable to obtain: 
\begin{equation}
  S =  S_t + S_U + S_{\text{ph}} + \int_0^{\beta} d\tau  \sum_n \frac{g_n}{\sqrt{N}}  X_n(\tau) \sum_{\ve{i}} \ve{i} \cdot \ve{\epsilon}_n \,  i \partial_{\tau} \left( \eta^{\dagger}_{\ve{i}}(\tau) \eta_{\ve{i}}(\tau) \right).
\end{equation} 
Local gauge invariance implies charge conservation such that 
\begin{equation}
  i \partial_{\tau} \left( \eta^{\dagger}_{\ve{i}}(\tau) \eta_{\ve{i}}(\tau) \right)  + \sum_{\ve{j}} J_{\ve{i},\ve{j}}(\tau) = 0.
\end{equation}
In the above, 
\begin{equation}
  J_{\ve{i},\ve{j}}(\tau) =  i \left( t_{\ve{i},\ve{j}} \eta^{\dagger}_{\ve{i}}(\tau) \eta_{\ve{j}}(\tau) - t_{\ve{j},\ve{i}} \eta^{\dagger}_{\ve{j}}(\tau) \eta_{\ve{i}}(\tau) \right),
\end{equation}
is the paramagnetic current operator. Therefore:
\begin{multline}
  \frac{1}{\sqrt{N}} \sum_{\ve{i}} \ve{i} \cdot \ve{\epsilon}_n \,  i \partial_{\tau} \left( \eta^{\dagger}_{\ve{i}}(\tau) \eta_{\ve{i}}(\tau) \right)   =
  - \frac{1}{\sqrt{N}} \sum_{\ve{i},\ve{j}} \ve{i} \cdot \ve{\epsilon}_n \, J_{\ve{i},\ve{j}}(\tau) \\ 
  =  -\frac{1}{2\sqrt{N}} \sum_{\ve{i},\ve{j}} \left( \ve{i} - \ve{j} \right) \cdot \ve{\epsilon}_n \, J_{\ve{i},\ve{j}}(\tau)   \equiv  J_{\epsilon_n}(\tau).
\end{multline}
In the above, we have used the fact that the current operator is antisymmetric under the exchange of the site indices. The action now reads: 
\begin{equation} 
  S =  S_t + S_U + S_{\text{ph}} + \int_0^{\beta} d\tau  \sum_n  g_n X_n(\tau) J_{\epsilon_n}(\tau).
\end{equation}

To proceed, we will insert source terms for the photon field, $\lambda_n(\tau)$,
\begin{equation} 
  S(\lambda) =  S_t + S_U + S_{\text{ph}} + \int_0^{\beta} d\tau  \sum_n ( g_n X_n(\tau) J_{\epsilon_n}(\tau) + \lambda_n(\tau) X_n(\tau)).
\end{equation}
Carrying out a   transformation  to  Matsubara  frequencies, 
\begin{gather} 
  \eta^{\dagger}(i\omega_m) = \frac{1}{\sqrt{\beta}} \int_0^{\beta} d\tau e^{i \omega_m \tau} \eta^{\dagger}(\tau), \\ 
  X_n(i\Omega_m) = \frac{1}{\sqrt{\beta}} \int_0^{\beta} d\tau e^{i \Omega_m \tau} X_n(\tau),
\end{gather}
where $\omega_m = (2m+1) \pi / \beta$ and $\Omega_m = 2 m \pi / \beta$ are fermionic and bosonic Matsubara frequencies, respectively,
we obtain: 
\begin{eqnarray}
  S(\lambda) && =  S_t + S_U +  \sum_{m>0, n} \Bigg[ \frac{1}{\Omega_n} \left( \Omega_m^2 + \Omega_n^2 \right) |X_n(i\Omega_m)|^2  +  \Bigg.\nonumber \\ 
  & &  \lambda_n^{\dagger}(i\Omega_m) X_n(i\Omega_m) + \lambda_n(i\Omega_m) X_n^{\dagger}(i\Omega_m) \\ 
  & &   \Bigg. +  g_n  X_n(i\Omega_m) J^{\dagger}_{\epsilon_n}(i\Omega_m) + g_n X_n^{\dagger}(i\Omega_m) J_{\epsilon_n}(i\Omega_m)  \Bigg].  \nonumber
\end{eqnarray}
In the above,  $J_{\epsilon_n}(i\Omega_m) =  \frac{1}{\sqrt{\beta}} \int_0^{\beta} d\tau e^{i \Omega_m \tau} J_{\epsilon_n}(\tau)$.
Gaussian integration over the photon fields yields: 
\begin{eqnarray}
  & & S(\lambda) = S_t + S_U - \sum_{m>0, n} \frac{\Omega_n}{ \Omega_m^2 + \Omega_n^2} \left[ g_n^2  J^{\dagger}_{\epsilon_n}(i\Omega_m) J_{\epsilon_n}(i\Omega_m)  \right.  \nonumber   \\ 
  & &  \left.     +  g_n \left(  \lambda_n^{\dagger}(i\Omega_m) J_{\epsilon_n}(i\Omega_m) + \text{H.c.} \right)  + |\lambda_n(i\Omega_m)|^{2} \right]. 
\label{eq:action_with_source}
\end{eqnarray}
In this integration step, we have neglected the boundary conditions of the $\eta$ fields that pick up an $X(\tau)$-dependent phase. Hence, strictly speaking, this step is 
only valid in the thermodynamic limit or on finite systems with open boundary conditions. 
Setting the source field to zero and carrying out a Fourier transform to imaginary time yields the effective action for the electrons of 
Eq.~\eqref{eq:effective_action}.
Differentiating with respect to the source fields allows us to obtain the photon propagator: 
\begin{multline}
  D(n, i \Omega_m) \equiv \langle X_n(i\Omega_m) X_n^{\dagger}(i\Omega_m) \rangle = \\
  D_0(n, i \Omega_m)  +  g_n^2  D_0^2(n, i \Omega_m) \left[ \langle J_{\epsilon_n}(i\Omega_m) J^{\dagger}_{\epsilon_n}(i\Omega_m) \rangle  - \langle J_{\epsilon_n}(i\Omega_m) \rangle \langle J^{\dagger}_{\epsilon_n}(i\Omega_m) \rangle\right],
\end{multline}
with $D_0(n, i \Omega_m) = \frac{ \Omega_n}{\Omega_m^2 + \Omega_n^2}$ the bare photon propagator of the \(n\)-th mode. The above establishes an exact relation between the photon propagator and the optical conductivity. We can now carry out the analytical continuation to 
real frequencies to obtain the exact relation between the photon spectral function and the optical conductivity. Specifically,  
Specifically, we can carry out the analytic continuation $i \Omega_m \to \omega + i 0^+$ and obtain the full photon propagator:
\begin{equation} \label{eq:D_Sigma_relation_EM}
  D(\omega) = \sum_n  D_0(n, \omega)  + i \omega \sum_n \left[ g_n^2 D_0^2(n, \omega) \right] \sigma^{\text{reg}}_{\epsilon_n,\epsilon_n}(\ve{q}=0,\omega)
\end{equation}
In the  above, the  regular part of the optical conductivity  reads: 
\begin{equation}
  \sigma^{\text{reg}}_{\epsilon_n,\epsilon_n} (\ve{q}=0,\omega) = \frac{1}{N}  \frac{ i \int_0^{\infty} dt e^{i \omega t} \langle [\hat{J}_{\epsilon_n}(0), \hat{J}_{\epsilon_n}(-t)] \rangle}{i (\omega + i0^+)}.
\end{equation}
Upon transforming back to the original fermion fields,  the  current operator  reads: 
\begin{eqnarray}
  \hat{J}_{\epsilon_n}  = & & -\frac{1}{2}\sum_{\ve{i},\ve{j}} i \left( \hat{c}^{\dagger}_{\ve{i}} e^{ i \frac{g}{\sqrt{N}} (\ve{j}  -\ve{i})\cdot \ve{\epsilon_n} \hat{X}}t_{\ve{i},\ve{j}} \hat{c}_{\ve{j}} - \hat{c}^{\dagger}_{\ve{j}} e^{i \frac{g}{\sqrt{N}} (\ve{i}-\ve{j})\cdot \ve{\epsilon_n} \hat{X}} t_{\ve{j},\ve{i}}  \hat{c}_{\ve{i}} \right)  \left( \ve{i} - \ve{j} \right) \cdot \ve{\epsilon_n}.
\end{eqnarray}
Eq.~\eqref{eq:D_Sigma_relation_EM} reveals why it is important to scale the interaction as $\sim 1/\sqrt{N}$ in the definition of the Hamiltonian. This scaling ensures that the photon number remains finite in the Mott phase where the conductivity is activated, as in the case of the empty cavity. Had we omitted the $1/\sqrt{N}$ scaling, the photon number in the cavity would diverge as soon as any material is present.

\subsection{Perturbative calculation of the electron-photon self-energy}
\label{subsec:self_energy}

In this  section, we compute the photon contribution to the electronic self-energy. For this we consider $U=0$ and perform an expansion in the light-matter coupling to the lowest order in perturbation theory. We will build on the path-integral formulation introduced in the previous section. 
Before proceeding, and in order to define the Brillouin zone,  we will assume that our tight-binding model is defined  on a Bravais lattice with a unit cell defined by \(\ve{R}\). Each orbital can  hence be assigned a unit  cell  and a  position in  the unit cell.  That is: 
\begin{equation}
  \ve{i} = \ve{R} + \ve{\delta}.
\end{equation}
Translation invariance  implies that $t_{\ve{i},\ve{j}} = t_{\ve{R}+\ve{\delta},\ve{R}'+\ve{\delta}'} = t_{\ve{\delta},\ve{\delta}'}(\ve{R}-\ve{R}')$.  Thereby, defining the Fourier transformation in space-time as:  
\begin{equation}
  \eta^{\dagger}_{\ve{k},\delta, n} =  \frac{1}{\sqrt{\beta N}} \int_0^{\beta} d\tau \sum_{\ve{R}} e^{i \omega_n \tau} e^{-i \ve{k} \cdot (\ve{R} + \ve{\delta})} \eta^{\dagger}_{\ve{R},\delta}(\tau),
\end{equation}
gives for the kinetic energy: 
\begin{equation}
  \sum_{\ve{i},\ve{j}}\eta^{\dagger}_{\ve{i}}(\tau) t_{\ve{i},\ve{j}} \eta_{\ve{j}}(\tau) = \sum_{\ve{k},\delta,\delta'} \eta^{\dagger}_{\ve{k},\delta}(\tau) t_{\delta,\delta'}(\ve{k}) \eta_{\ve{k},\delta'}(\tau)
\end{equation}
with $t_{\delta,\delta'}(\ve{k}) = \sum_{\Delta \ve{R}} e^{-i \ve{k} \cdot (\Delta \ve{R} - \ve{\delta} + \ve{\delta}')} t_{\ve{\delta},\ve{\delta}'}(\Delta \ve{R})$.  The current operator reads:
\begin{equation}
  J_{\epsilon_n}(\tau) = \sum_{\ve{k},\delta,\delta'} \eta^{\dagger}_{\ve{k},\delta}(\tau) j_{\delta,\delta'}(\ve{k})  
\eta^{\phantom{\dagger}}_{\ve{k},\delta'}(\tau)
\end{equation}
with the current matrix
\begin{multline}
  j_{\delta,\delta'}(\ve{k}) = -\frac{1}{2} \sum_{\Delta \ve{R}} \Big( 
  i t_{\ve{\delta},\ve{\delta}'}(\Delta \ve{R}) e^{-i \ve{k} \cdot (\Delta \ve{R} - \ve{\delta} + \ve{\delta}')} \left( \Delta \ve{R} + \ve{\delta} - \ve{\delta}' \right) \cdot \ve{\epsilon_n} - \\
  \overline{t_{\ve{\delta}',\ve{\delta}}(\Delta \ve{R})} e^{i \ve{k} \cdot (\Delta \ve{R} + \ve{\delta}' - \ve{\delta})} \left( \Delta \ve{R} - \ve{\delta}' + \ve{\delta} \right) \cdot \ve{\epsilon_n} \Big) 
\end{multline}
With this  notation,  the partition function reads: 
\begin{equation}
  Z = \int D\left\{ \eta^{\dagger}_{\ve{k},\delta}(\tau), \eta_{\ve{k},\delta}(\tau), X_n(\tau) \right\} e^{-S_{\text{ep}}[\eta^{\dagger}_{\ve{k},\delta}(\tau), \eta_{\ve{k},\delta}(\tau), X_n(\tau)] -  S_{p} [X_n(\tau)]}
\end{equation}
with the electron-photon action 
\begin{equation}
  S_{\text{e-ph}} = \int_0^{\beta} d\tau \left[ \sum_{\ve{k}} \eta^{\dagger}_{\ve{k}}(\tau) (\partial_{\tau} + t(\ve{k})) \eta_{\ve{k}}(\tau)  - \sum_n \frac{g_N}{\sqrt{N}} X_n(\tau) \eta^{\dagger}_{\ve{k}}(\tau) j(\ve{k}) \eta_{\ve{k}}(\tau) \right]
\end{equation}
and photon action
\begin{equation}
  S_{\text{ph}} =  \frac{1}{2} \int_0^{\beta} d\tau \sum_n \left[ \frac{1}{\Omega_n}  \dot{X}_n^2(\tau) + \Omega_n X_n^2(\tau) \right].
\end{equation}
As is apparent, the action is diagonal in momentum, reflecting the fact that the photon field does not carry any momentum. We can now integrate out the photon field to obtain the effective action for the  electrons,
\begin{equation}
  Z = \prod_{\ve{k}} \int D\left\{ \eta^{\dagger}_{\ve{k},\delta}, \eta_{\ve{k},\delta} \right\} e^{-S_{\text{e}}(\ve{k},\eta^{\dagger}_{\ve{k},\delta}, \eta_{\ve{k},\delta})}
\end{equation}
with 
\begin{multline}
  S_{\text{e}}(\ve{k})   =  \sum_{i\omega_m} \eta^{\dagger} (\ve{k},i\omega_m) \left( i \omega_m + t(\ve{k}) \right) \eta(\ve{k},i\omega_m)  - \\
  \sum_n \frac{g_n^2}{2} \sum_{i\Omega_n} D_0(n, i\Omega_n) J^{\dagger}_{\epsilon_n}(\ve{k},i\Omega_n) J_{\epsilon_n}(\ve{k},i\Omega_n)
\end{multline}
and 
\begin{equation}
  J_{\epsilon_n}(\ve{k},i\Omega_n)   =  \frac{1}{\sqrt{N \beta}} \sum_{i \omega_m} \eta^{\dagger}(\ve{k},i\omega_m) j(\ve{k}) \eta(\ve{k},i\omega_m + i \Omega_n).
\end{equation}
The  above  equation reveals that for a s single  momentum $\ve{k}$  the  interaction mediated by the photons scales as $g_n^2/N$.  Since  $g_n$  is a   size-independent constant, the effective interaction  vanishes in the thermodynamic limit.
We can see this more explicitly by computing the electron self-energy in lowest-order perturbation theory. We obtain: 
\begin{equation} 
  \Sigma_{\text{e-ph}}(\ve{k},i\omega_m) =   \sum_n \frac{g_n^2}{N}  j(\ve{k}) \frac{1}{\beta}\sum_{i\Omega_n} D_0(n, i\Omega_n)  G_0(\ve{k},i\omega_m + i \Omega_n) j(\ve{k})   
\end{equation}
with the  bare electron Green's function $G_0(\ve{k},i\omega_m) = 1/(i \omega_m + t(\ve{k}))$.  As the photons do not couple to the Hubbard interaction, the self-energy due to the Hubbard interaction will only renormalise the single-particle Green's function. Thus, the full interacting Green's function is given by \(G(\vec{k}, i\omega_m) = 1/(i\omega_m + t(\vec{k}) - \Sigma_{\text{e-ph}}(\ve{k},i\omega_m) - \Sigma_{U}(\ve{k},i\omega_m))\), where \(\Sigma_{U}\) is the self-energy due to the Hubbard interaction.

\subsection{Auxiliary-field quantum Monte Carlo formulation}
\label{subsec:af-qmc_formulation}

In this section we formulate the auxiliary-field quantum Monte Carlo (AF-QMC) method for the light-matter Hamiltonian, Eq.~\eqref{eq:hamiltonian} in the main text. For this we write the Hamiltonian in the path integral representation and integrate out the fermions to get an effective bosonic action we can sample with Monte Carlo.

In order to write down the action of the Hamiltonian of the photon, we first need to define the basis vectors of the $\hat{X}$ and $\hat{P}$ operators, $[\hat{X}, \hat{P}] = i$, and their overlap. We define 
\begin{equation}
  \hat{X} \ket{x} = x \ket{x}, \quad \hat{P} \ket{p} = p \ket{p}
\end{equation}
where $x, p \in \mathbb{R}$ and $\braket{x}{x^\prime} = \delta(x-x^\prime)$ and the same for $p$. The representation of $\hat{P}$ in the $\hat{X}$ basis is \(\hat{P} \to - i \frac{\partial}{\partial x}\).
The overlap between the eigenstates of \(\hat X\) and \(\hat P\) is obtained from
\begin{align}
  \bra{x} \hat{P} \ket{p} &= p \braket{x}{p} \nonumber \\
  -i \frac{\partial}{\partial x} \braket{x}{p} &= p \braket{x}{p} \nonumber \\
  \braket{x}{p} &= A e^{ipx},
\end{align}
where $A$ is a constant. We can find $A$ by the completeness relation of the eigenstates \(1 = \int dx \ketbra{x}{x} = \int \frac{dp}{2\pi} \ketbra{p}{p}\). We start with 
\begin{equation}
  \braket{p}{p^\prime} = \int dx \braket{p}{x}\braket{x}{p^\prime} = \left|A\right|^2 \int dx  e^{- ix\left(p - p^\prime\right)} = \left|A\right|^2 2 \pi \delta(p - p^\prime) \overset{!}{=} 2\pi \delta(p - p^\prime),
\end{equation}
so $A = 1$ and \(\braket{x}{p} = e^{ipx} \). Now we can calculate the path integral of the Hamiltonian of the photon mode. We start with a Trotter discretisation of the imaginary-time propagator into  $L_\tau$ time slices of size $\Delta\tau$ ($L_\tau \Delta\tau = \beta$) and perform a path integral using the eigenstates of the $\hat{X}$ operator:
\begin{equation}
  Z_{\text{ph}} = \text{Tr} \left[ e^{-\beta\hat{H}_{\text{ph}}} \right] = \int \D{x} \prod_{\tau = 1}^{L_\tau} \bra{x_{\tau+1}} e^{-\Delta\tau \hat{H}_{\text{ph}}} \ket{x_\tau},
\end{equation}
with the condition that $\ket{x_1} = \ket{x_{L_\tau + 1}}$, \(\hat{H}_{\text{ph}} = \frac{\Omega}{2}(\hat{P}^2 + \hat{X}^2)\) and \(\beta = 1/T\). We now focus on computing the matrix element present in the path integral, 
\begin{align}
  \bra{x_{\tau+1}} & e^{-\Delta\tau \hat{H}_{\text{ph}}} \ket{x_\tau} = \int dp \braket{x_{\tau+1}}{p} \bra{p} e^{-\Delta\tau \Omega \hat{P}^2 / 2} \ket{x_\tau} e^{-\Delta\tau \Omega x_\tau^2 / 2} \nonumber \\
  &\sim e^{-\Delta\tau \Omega x_\tau^2 / 2} \int dp  e^{i p \left(x_{\tau+1} - x_\tau\right) - \Delta\tau \Omega p^2 / 2} \sim e^{-\Delta\tau \Omega x_\tau^2 / 2} e^{-\frac{\left(x_{\tau+1}-x_\tau\right)^2}{2\Delta\tau\Omega}}.
\end{align}
So the partition function becomes \(Z_{\text{ph}} = \int \D{x} e^{-S_{\text{ph}}}\), where 
\begin{equation}
  S_{\text{ph}} = \sum_{\tau=1}^{L_\tau} \left(\frac{\left(x_{\tau+1}-x_\tau\right)^2}{2 \Delta\tau \Omega} + \Delta\tau \frac{\Omega}{2} x_\tau^2 \right)
\end{equation}
is the action for the photon mode.

Now we focus on the Hamiltonian with electronic degrees of freedom, equation \eqref{eq:hamiltonian}. After an imaginary-time discretisation, the partition function is given by 
\begin{equation}
  Z = \text{Tr}\left[e^{-\Delta\tau (\hat{H}_t + \hat{H}_U + \hat{H}_{\text{ph}})}\right]^{L_\tau},
\end{equation}
where 
\begin{gather}
  \hat{H}_t = -t \sum_{i, \delta, \sigma} \left(\cd_{i, \sigma} e^{i g_\delta \hat{X}} \hat{c}_{i+\delta, \sigma} + \text{h.c.}\right) = - t \sum_{i,\delta} \hat{K}_{i,\delta}(\hat{X}), \\
  \hat{H}_U = \frac{U}{2} \sum_i (\n_i - 1)^2 = \frac{U}{2} \sum_i \hat{U}_i^2,
\end{gather}
and \(\hat{H}_{\text{ph}}\) is the photon Hamiltonian. To preserve the Hermiticity of the Hamiltonian we use a symmetric Trotter decomposition 
\begin{equation}
  e^{-\Delta\tau (\hat{H}_t + \hat{H}_U + \hat{H}_{\text{ph}})} = \left(\prod_{b = 1}^{N_b} e^{-\frac{\Delta\tau}{2} \hat{H}_{t,b}} \right) e^{-\Delta\tau \hat{H}_U} e^{-\Delta\tau \hat{H}_{\text{ph}}} \left(\prod_{b = N_b}^{1} e^{-\frac{\Delta\tau}{2} \hat{H}_{t,b}} \right) + \mathcal{O}((\Delta\tau)^3),
\end{equation}
where \(N_b\) is the number of bonds. The Hubbard interaction is decoupled with a discrete Hubbard-Stratonovich (HS) transformation:
\begin{equation}
  e^{- \Delta\tau \frac{U}{2} (\hat{n}_i - 1)^2} = \frac{1}{4} \sum_{l = \pm 1, \pm 2} \gamma(l) e^{\sqrt{- U\Delta\tau/2} \eta(l) (\n_i - 1)} + \mathcal{O}((\Delta\tau)^4),
\end{equation}
where 
\begin{gather}
  \gamma(\pm 1) = 1 + \sqrt{6}/3, \quad \eta(\pm 1) = \pm \sqrt{2(3 - \sqrt{6})}, \\
  \gamma(\pm 2) = 1 - \sqrt{6}/3, \quad \eta(\pm 2) = \pm \sqrt{2(3 + \sqrt{6})}.
\end{gather}
Since the Trotter decomposition introduces a systematic error of \(\mathcal{O}((\Delta\tau)^3)\), we can consider the discrete HS transformation exact. Integrating out the fermions, we arrive at
\begin{equation}
  Z = \sum_{\left\{l_{i,\tau}\right\}} \left(\prod_{i,\tau} \frac{\gamma(l_{i,\tau})}{4}\right) \int \D{x}  e^{-S_{\text{ph}}} \det\left(1 + B(\beta, 0)\right)
\end{equation}
where the single-particle propagation matrix is given by
\begin{equation} 
  B(\tau_2, \tau_1) = \prod_{\tau=\tau_1 + \Delta\tau}^{\tau_2} \left(\prod_{b=1}^{N_b} e^{\frac{\Delta\tau}{2} t K_{i,\delta}(x_\tau)}\right) \left(\prod_{i=1}^N e^{\sqrt{-\Delta\tau U / 2} \eta_{i,\tau} U_i} \right) \left(\prod_{b=N_b}^{1} e^{\frac{\Delta\tau}{2} t K_{i,\delta}(x_\tau)}\right)
\end{equation}
This action can be sampled using the auxiliary-field quantum Monte Carlo method where the configuration space is the set \(C = \{l_{i,\tau}; x_\tau \}\) with a discrete HS field per site and time slice and one continuous photon field per time slice. To update the configuration we use local updates for \(l_{i,\tau}\) and a global update in time for the photon field where the new field is taken from a box distribution \(x^\prime_\tau = x_\tau + (U(0,1) - 1/2)\). Here \(U(0,1)\) is a uniformly distributed random number between 0 and 1.

Despite the complex Peierls phase factors in the hopping matrix, the AF-QMC formulation is free of the negative sign-problem at half-filling. To demonstrate this, we perform a PH transformation on the down-spin sector:
\begin{equation}
  \hat{d}_{i,\uparrow} = \c_{i,\uparrow}, \qquad \hat{d}_{i,\downarrow}= (-1)^i \c_{i,\downarrow}^{\dagger},
\end{equation}
where \((-1)^i\) is \(+1\) on sublattice \(A\) and \(-1\) on sublattice \(B\). Since hopping connects opposite sublattices, this transformation makes the down-spin hopping matrix the complex conjugate of the up-spin hopping matrix for any photon configuration. Furthermore, defining $\phi_{i,\tau} = \sqrt{U\Delta\tau/2}\,\eta(l_{i,\tau}) \in \mathbb{R}$, the HS factor becomes
\begin{equation}
  e^{i\phi_{i,\tau}(\hat{n}_i-1)} = e^{i\phi_{i,\tau} (\hat{d}_{i,\uparrow}^{\dagger}\hat{d}_{i,\uparrow} - \hat{d}_{i,\downarrow}^{\dagger}\hat{d}_{i,\downarrow})}.
\end{equation}
Thus, the two transformed sectors have complex conjugate imaginary-time propagators, and the relation \(B_{\downarrow}(C)=B_{\uparrow}(C)^*\) holds. The Monte Carlo weight therefore is given by 
\begin{equation}
  W(C)= e^{-S_{\mathrm{ph}}[x]} \left(\prod_{i,\tau}\frac{\gamma(l_{i,\tau})}{4}\right) \left|\det\left[1+B_{\uparrow}(C)\right]\right|^2 \geq 0.
\end{equation}
Since the photon action is real and all HS coefficients $\gamma(l)$ are positive, every configuration has a nonnegative weight \cite{wu_sufficient_2005}. This argument holds for arbitrary electron-photon coupling and relies on half filling and the spin-independent, bipartite hopping matrix.

\begin{figure*}[t]
  \centering
  \begin{tabular}{cc}
    (a) & (b) \\
    \includegraphics[width=0.45\textwidth]{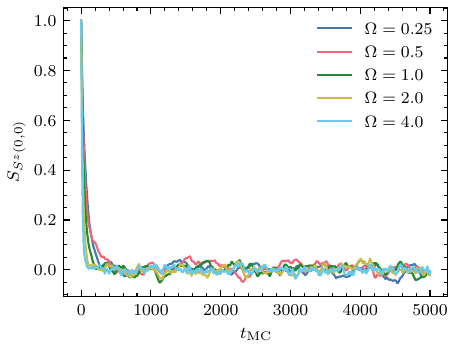} & 
    \includegraphics[width=0.45\textwidth]{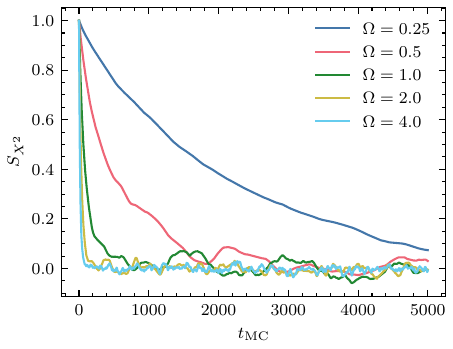}\\
    (c) & (d) \\
    \includegraphics[width=0.45\textwidth]{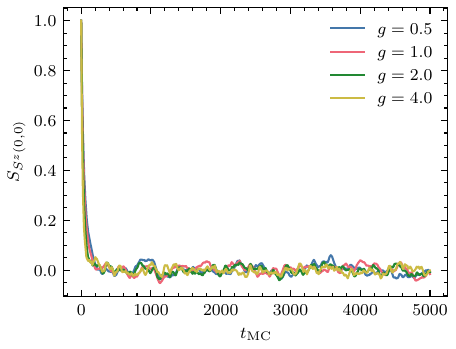} & 
    \includegraphics[width=0.45\textwidth]{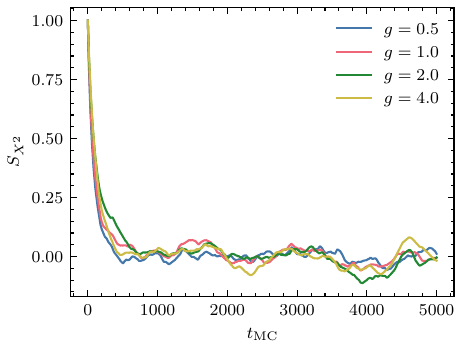}  
  \end{tabular}
  \caption{
    \textbf{Monte Carlo autocorrelation functions of spin and photon observables.}
    \textbf{a,b}, Normalised autocorrelation functions of the uniform spin structure factor and of the squared cavity coordinate \(\hat{X}^{2}\), respectively, versus Monte Carlo time \(t_{\mathrm{MC}}\) at \(g = 1\). Different curves denote \(\Omega/t =0.25, 0.5, 1, 2\) and \(4\).
    \textbf{c,d}, Corresponding autocorrelation functions at \(\Omega/t =1\), where different curves denote \(g=0.5, 1, 2\) and \(4\). 
    One unit of \(t_{\mathrm{MC}}\) denotes a sweep in which every field on the Euclidean space-time lattice is visited once. Here \(U/t = 0\), \(L = 6\), \(\beta t = L\).
  }
  \label{fig:autocorrelation}
\end{figure*}

Fig.~\ref{fig:autocorrelation} shows the autocorrelation functions for the uniform (\(\vec{q} = (0, 0)\)) spin structure factor and photon field \(\hat{X}^2\) for different values of \(\Omega\), Figs.~\ref{fig:autocorrelation}(a) and \ref{fig:autocorrelation}(b), and \(g\), Figs.~\ref{fig:autocorrelation}(c) and \ref{fig:autocorrelation}(d). The autocorrelations in the spin structure factor are largely unchanged by tuning \(\Omega\) and \(g\). On the other hand, the autocorrelation times of the photon field increase when \(\Omega \to 0\) and remain unchanged as a function of \(g\). This increase at lower frequencies is expected, since the photon field dynamics become slower as \(\Omega\) decreases, making efficient sampling with a box distribution more challenging. However, for \(\Omega = 1\) the autocorrelation times of the photon field are in the order \(10^2\), indicating that our sampling scheme is still quite efficient in this regime.

\section*{Acknowledgements}

We thank discussions with Z. Bacciconi, D. Baykusheva,  I. Herbut,  W. Kennedy, A. P\'alffy-Buss, F. Piazza, A. Reingruber, M. Sentef,  J. Silva and M. Ulybyshev.
We  gratefully acknowledge the Gauss Centre for Supercomputing e.V. for funding this project by providing computing time on the GCS Supercomputer SUPERMUC-NG at Leibniz Supercomputing,   (project number pn73xu) as  well  as  the scientific support and HPC resources provided by the Erlangen National High Performance Computing Center (NHR@FAU) of the Friedrich-Alexander-Universit\"at Erlangen-N\"urnberg (FAU) under the NHR project b133ae. NHR funding is provided by federal and Bavarian state authorities. NHR@FAU hardware is partially funded by the German Research Foundation (DFG) -- 440719683. 
J.C.I. thanks the DFG for financial support under the AS 120/19-1 grant (Project number, 530989922).
F.F.A.\ acknowledge financial support  from the DFG under the grant AS 120/16-1 (Project number 493886309) that is part of the collaborative research project SFB Q-M\&S funded by the Austrian Science Fund (FWF) F 86 as well as from the W\"urzburg-Dresden Cluster of Excellence {\it ctd.qmat} (EXC 2147, Project No.~390858490).
N.C.C.~acknowledges support from the Brazilian funding agencies CNPq [Grant No.~308130/2025-1], CAPES, FAPERJ [Grants No.~E-26/200.258/2023 and E-26/210.592/2025], and Serrapilheira Institute [Grant No.~R-2502-52037]. N.C.C.~also acknowledges support from Alexander von Humboldt Foundation.

\section*{Author contributions}

J.C.I. implemented the AF-QMC method and performed the numerical work. F.F.A. and J.C.I.  performed the analytical work and wrote the manuscript. N.C.C. was involved in the conceptualisation of the project and manuscript writing. 
All the authors contributed to discussing the results.

\section*{Competing interests}

The authors declare no competing interests.

\section*{Data availability}

The numerical data presented in this work is available at request to the authors.

\section*{Code availability}

The AF-QMC and SAC codes are related to the publicly available ALF package \cite{assaad_alf_2025}. The specific Hamiltonian file used for the light-matter Hamiltonian is available upon request from the authors.

\bibliography{joao.bib, fakher.bib}

\clearpage

\section{Supplementary information} 
\label{sec:sm}

\subsection{\(C_{3v}\) point group symmetry}

In this section, we show that the electron-photon coupling breaks all of the point group \(C_{3v}\) symmetries of the honeycomb lattice. To this end, we consider the coupling 
\begin{equation}
  g_\delta = \frac{g}{\sqrt{N}} \vec{\epsilon} \cdot \vec{n}_{\delta},
\end{equation}
where we define the lattice vectors as \(\vec{a}_1 = (1, 0)\), \(\vec{a}_2 = \frac{1}{2}(1, \sqrt{3})\) and the nearest-neighbour (NN) vectors as
\begin{equation}
  \vec{n}_1 = \frac{1}{3} (0, \sqrt{3}),\quad \vec{n}_2 = \frac{1}{6} (-3, -\sqrt{3}), \quad \vec{n}_3 = \frac{1}{6} (3, -\sqrt{3}).
\end{equation}
The point group \(C_{3v}\) has 6 elements: 3 \(2n\pi/3\) rotations \(R^n_3\) (\(n = 1, 2, 3\)); 3 mirror symmetries \(\sigma_1: (x, y) \to (-x, y)\), \(\sigma_2 = R^2_3\sigma_1\) and \(\sigma_3 = R_3 \sigma_1\). Under these operations, the NN vectors transform as 
\begin{gather}
  R_3^n \vec{n}_{\delta} \to \vec{n}_{\mod(\delta + n, 3)}, \\
  \sigma_n \vec{n}_{\delta} \to \delta_{n,\delta} \vec{n}_\delta + (1 - \delta_{n,\delta}) \vec{n}_{\delta^\prime \neq n}.
\end{gather}
It is then clear that under one rotation and for any choice of \(\vec{\epsilon}\), \(R_3^n g_\delta \to g_{\mod(\delta + n, 3)}\). Furthermore, \(g_\delta\) is not invariant under a mirror transformation unless \(\vec{\epsilon}\) lies on the mirror axis. Since we choose \(\vec{\epsilon} = (1, 0)\), all mirror symmetries are also broken. Breaking the lattice point group symmetries implies that the Dirac cones may meander and cause a nematic transition \cite{schwab_nematic_2022}. We calculate the photon-induced self-energy and show that it vanishes in the thermodynamic limit, thus prohibiting any meandering of the Dirac cones.

\subsection{Equations of motion}

In this section we present the equation of motion for the photon operator \(\hat{a}\): \(\partial_\tau \hat{a}(\tau) = e^{\tau \hat{H}} [\hat{H}, \hat{a}] e^{-\tau \hat{H}}\), 
\begin{equation} \label{eq:equation_motion_photon} 
  \partial_\tau \hat{a}(\tau) = \sum_{\delta} \frac{g_{\delta}}{\sqrt{2}}  \big[ \cos(g_\delta (\a(\tau) + \ad(\tau))/\sqrt{2}) \hat{J}_{\vec{q} = 0, \delta}(\tau) - \sin(g_\delta (\a(\tau) + \ad(\tau))/\sqrt{2}) \hat{K}_{\vec{q} = 0, \delta}(\tau) \big] - \Omega \a(\tau),
\end{equation}
where \(\hat{K}_{\vec{q}, \delta}(\tau) = \sum_{\vec{k}, \sigma} \hat{\psi}^\dagger_{\vec{k}+\vec{q}, \sigma}(\tau) \left( \vec{n}_{\vec{k}, \delta} \cdot \vec{\tau} \right) \psi_{\vec{k}, \sigma}(\tau) \) and \(\hat{J}_{\vec{q}, \delta}(\tau) = \sum_{\vec{k}, \sigma} \hat{\psi}^\dagger_{\vec{k} + \vec{q}, \sigma}(\tau) \left( \vec{v}_{\vec{k}, \delta} \cdot \vec{\tau} \right) \psi_{\vec{k}, \sigma}(\tau) \) are the kinetic-energy and current operators, respectively. The spinor is defined as  \(\hat{\psi}_{\vec{k}, \sigma} = (\hat{c}_{\vec{k}, A, \sigma} \hat{c}_{\vec{k}, B, \sigma})^T\) and the matrix elements are given by \(\vec{n}_{\vec{k},\delta} = (\text{Re} \gamma_{\vec{k}, \delta}, -\text{Im} \gamma_{\vec{k}, \delta}, 0 )\), with \(\gamma_{\vec{k}, \delta} = - t e^{- i \vec{k} \cdot \vec{n}_\delta}\) and \(\vec{v}_{\vec{k}, \delta} = \partial_{\vec{k}} \vec{n}_{\vec{k}, \delta}\). Equation \eqref{eq:equation_motion_photon} shows that the photons couple exclusively to fermionic bilinears of the form \(\hat{\psi}^\dagger_{\vec{k},\sigma}(\tau) \tau^i \hat{\psi}_{\vec{k}, \sigma}(\tau)\), corresponding to particle-hole (PH) excitations with momentum transfer \(\vec{q}=0\) and total spin \(S=0\). This is consistent with the approach developed in the Methods subsection on the path-integral formulation in the main text, where we show that electronic contributions to the photon spectral function are the dynamic current-current correlations. By this analysis we reach the same conclusion that the photons are irrelevant at the Gross-Neveu critical point. That is, there is a mismatch between the quantum numbers of the critical modes (magnons) and those of the PH excitations to which the photons couple.

\subsection{Electronic spectral function}

\begin{figure*}[t]
  \centering
  \begin{tabular}{cc}
    (a) & (b) \\ 
    \includegraphics[width=0.4\textwidth]{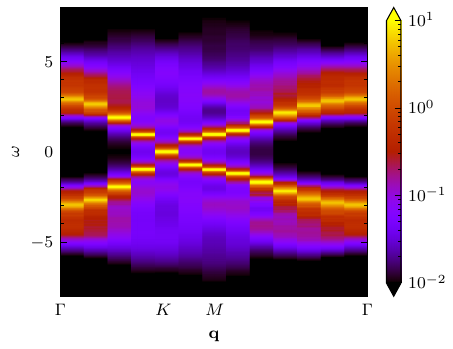} & 
    \includegraphics[width=0.4\textwidth]{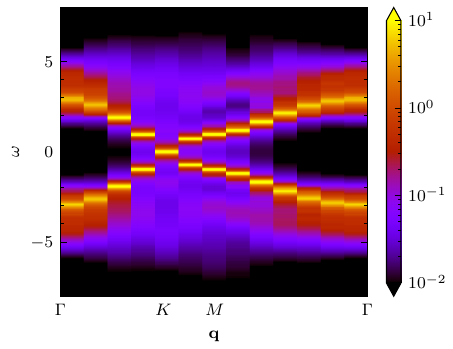} \\
    (c) & (d) \\
    \includegraphics[width=0.4\textwidth]{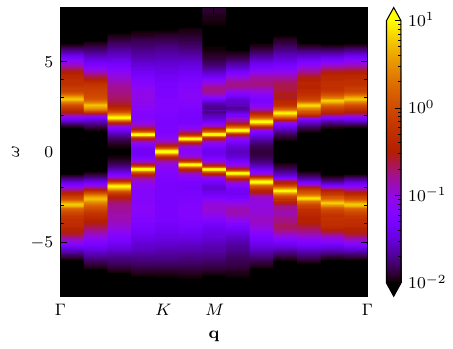} & 
    \includegraphics[width=0.4\textwidth]{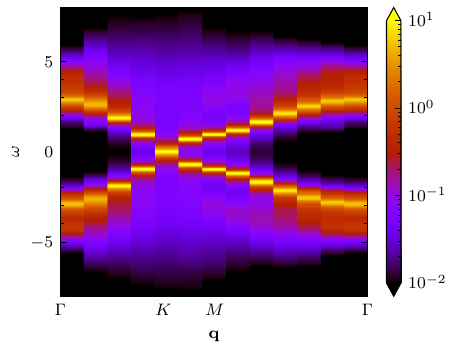}  
  \end{tabular}
  \caption{Electronic spectral function \(A(\vec{k}, \omega)\) for (a) \(g = 0.5\), (b) \(g = 1\), (c) \(g = 2\) and (d) \(g = 4\), in the DSM phase at \(U = 2\). Here \(L = 12\) and \(\beta = L\).}
  \label{fig:spectral_fermion_gs_DSM}
\end{figure*}

\begin{figure*}[t]
  \centering
  \begin{tabular}{cc}
    (a) & (b) \\
    \includegraphics[width=0.4\textwidth]{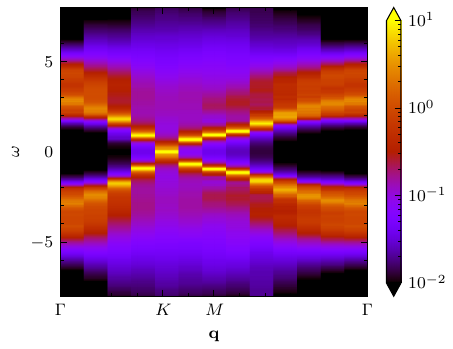} & 
    \includegraphics[width=0.4\textwidth]{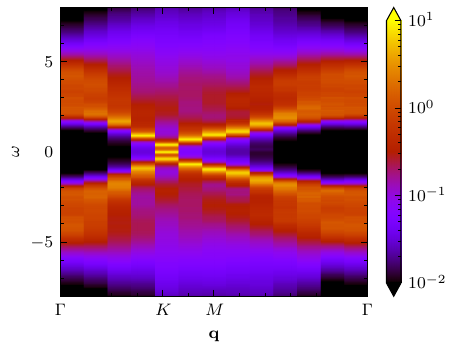} \\
    (c) & (d) \\
    \includegraphics[width=0.4\textwidth]{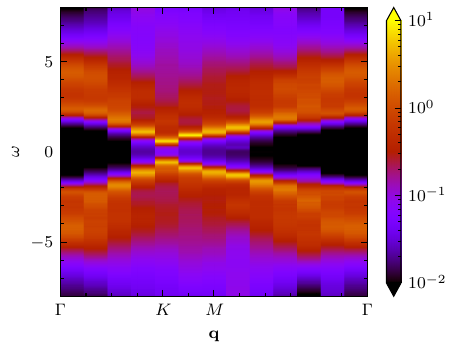} & 
    \includegraphics[width=0.4\textwidth]{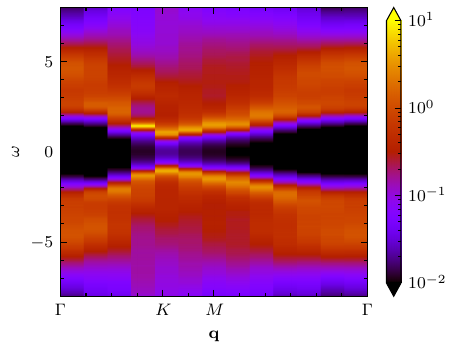}  
  \end{tabular}
  \caption{Electronic spectral function \(A(\vec{k}, \omega)\) for (a) \(U = 3\), (b) \(U = 4\), (c) \(U = 5\) and (d) \(U = 6\), for \(g = 4\). Here \(L = 12\) and \(\beta = L\).}
  \label{fig:spectral_fermion_gs_AFM}
\end{figure*}

In this section we show the electronic spectral function \(A(\vec{k}, \omega)\) obtained by applying the SAC method to the imaginary-time data for the single-particle Green's function \(G(\vec{k}, \tau)\). Figs. \ref{fig:spectral_fermion_gs_DSM}(a)-(d) present \(A(\vec{k}, \omega)\) for \(g = 0.5, 1, 2, 4\), respectively, in the DSM phase at \(U = 2\). As a function of \(g\), the electronic spectral function is largely unchanged, i.e. no new modes emerge. Increasing \(g\), we can see a broadening of the spectral width across different momenta. This is consistent with a decrease in the quasiparticle lifetime due to the emergence of polaritonic modes. Figs. \ref{fig:spectral_fermion_gs_AFM}(a)-(d) present \(A(\vec{k}, \omega)\) for \(U = 3, 4, 5, 6\), respectively, at strong electron-photon coupling \(g = 4\). The AFM mass gap develops at \(U \approx 4\), as evidenced by Fig. \ref{fig:spectral_fermion_gs_AFM}(b).

\subsection{Density and spin spectral functions}

\begin{figure*}[t]
  \centering
  \begin{tabular}{cccc}
    (a) & (b) \\
    \includegraphics[width=0.4\textwidth]{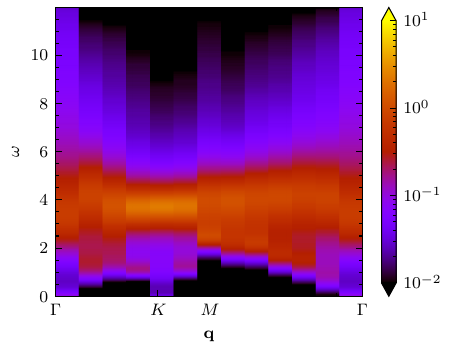} & 
    \includegraphics[width=0.4\textwidth]{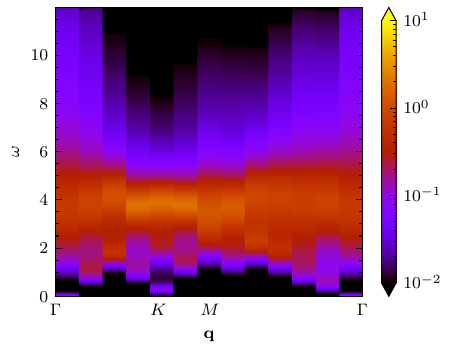} \\
    (c) & (d) \\
    \includegraphics[width=0.4\textwidth]{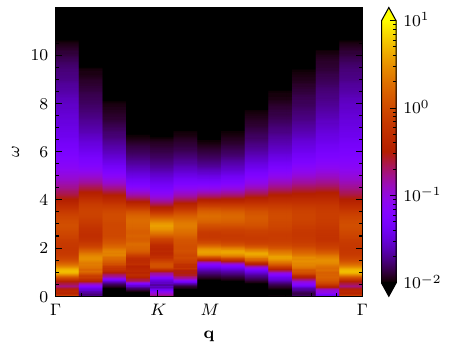} & 
    \includegraphics[width=0.4\textwidth]{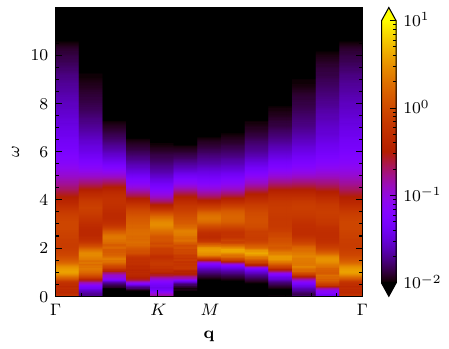}  
  \end{tabular}
  \caption{Momentum resolved density \(S^{\text{den}}(\vec{q}, \omega)\) (a)-(b) and spin \(S^{\text{spin}}(\vec{q}, \omega)\) (c)-(d) spectral function for \(g = 0.5\) and \(g = 4\), respectively, at \(U = 2\). Here \(L = 12\) and \(\beta = L\).}
  \label{fig:spectral_den_spin_gs}
\end{figure*}

In this section we show the density and spin spectral functions \(S^{\text{den}}(\vec{q}, \omega)\) and \(S^{\text{spin}}(\vec{q}, \omega)\), respectively, obtained by applying the SAC method to the imaginary-time data for the density and spin correlations. The imaginary-time density correlation function is defined as \(S^{\text{den}}(\vec{q}, \tau) = \langle \hat{n}_{\vec{q}}(\tau) \hat{n}_{-\vec{q}} (0)\rangle\), where \(\hat{n}_{\vec{q}}\) is the number operator, and the imaginary-time spin correlation function is given by \(S^{\text{spin}}(\vec{q}, \tau) = \langle \vec{\hat{S}}_{\vec{q}}(\tau) \cdot \vec{\hat{S}}_{-\vec{q}} (0)\rangle\), where \(\vec{\hat{S}}_{\vec{q}}\) is the spin operator. Figures \ref{fig:spectral_den_spin_gs}(a)--(b) show the density spectral function, and Figs. \ref{fig:spectral_den_spin_gs}(c)--(d) show the spin spectral function, at \(U=2\). Within each pair, \(g=0.5\) and \(4\), respectively. Consistent with the weak influence of the cavity coupling on collective charge and spin excitations, the spectral features are largely unchanged upon increasing the photon coupling strength. 

\end{document}